\RequirePackage{fixltx2e}
\documentclass[floatfix,twocolumn,a4paper]{revtex4-1}

\usepackage{amsmath}

\usepackage{mathtools}

\usepackage{amssymb}
\usepackage{amsfonts}

\usepackage{relsize}
\usepackage{graphicx}
\usepackage{placeins}

\usepackage[none]{hyphenat}
\usepackage{graphicx, subfigure}

\usepackage{todonotes}

\renewcommand{\d}[1]{\ensuremath{\operatorname{d}\!{#1}}}

\newcommand{\mean}[1]{\left<{#1}\right>}

\newcommand{\ER}{Erd\H{o}s-R\'enyi }

\begin{document}
\title{Meta-food-chains as a many-layer epidemic process on networks} 
\author{Edmund Barter}
\author{Thilo Gross}
\affiliation{University of Bristol, Department of Engineering Mathematics and Bristol Centre for
Complexity Sciences, Bristol, UK}
\begin{abstract}
Notable recent works have focused on the multi-layer properties of coevolving diseases. We point out that very similar systems play an important role in population ecology. Specifically we study a meta food-web model that was recently proposed by Pillai et al. This model describes a network of species connected by feeding interactions, which spread over 
a network of spatial patches. Focusing on the essential case, where the network of feeding interactions is a chain, we develop an analytical approach for the computation of the degree distributions of colonized spatial patches for the different species in the chain. This framework allows us to address ecologically relevant questions. Considering configuration model ensembles of spatial networks, we  find that there is an upper bound for the fraction of patches that a given species can occupy, which depends only on the networks mean degree. For a given mean degree there is then an optimal degree distribution that comes closest to the upper bound. Notably scale-free degree distributions perform worse than more homogeneous degree distributions if the mean degree is sufficiently high. Because species experience the underlying network differently the optimal degree distribution for one particular species is generally not the optimal distribution for the other species in the same food web. These results are of interest for conservation ecology, where, for instance, the task of selecting areas of old-growth forest to preserve in an agricultural landscape, amounts to the design of a patch network. 
\end{abstract}
\maketitle

\section{Introduction}
Over the past decade mathematical models have continued to provide insights into complex systems. Much of this work was grounded in the exploration of simple models from many areas, including ecology and epidemiology \cite{Newman1999,More200,Newman2002}. As the theoretical tools for the analysis of networks have been refined, the focus of current work has shifted to extend the range of systems that can be treated as networks. While earlier network models focused on structural properties of static systems \cite{DeSollaPrice1965,Callaway2000}, recent advances focus on dynamic and properties of time varying networks\cite{Gross2008, Holme2012}. Similarly the information encoded in networks has evolved from simple, often binary, variables to structures in which nodes and links can assume complex states in what are often described as multi-layer or multiplex networks \cite{Gomez2013,DeDomenico2013,Boccaletti2014,Kivela2014,Lee2015}. 

Theoretical progress in the analysis of complex networks both enables and requires a progression towards more complex models. For example one extension to simple epidemic models are coinfection models, which describe the simultaneous and interdependent spreading of two diseases \cite{Newman2005,Ahn2006b,Funk2010,Karrer2011,Zhu2013a,Newman2013,Zhao2014}. While these models have received some attention, very similar challenges encountered in ecology remain largely unexplored.

In ecology it is widely recognized that our environment is not evenly distributed. Typical landscapes are broken into distinct habitat patches. Examples include patches of forest remaining in an agricultural landscape, islands, in an archipelago, systems of lakes, or parks in a city,. A network representation of the environment can be constructed by using nodes to represent the discrete patches of habitat with links between pairs of patches between which a species can spread.

Another type of network that is considered in ecology are food webs, the networks of who eats who. In a food web the nodes represent populations of different species and directed links represent trophic (i.e. predator-prey) interactions.

An emerging topic in the ecological literature are so-called meta-foodwebs \cite{Pillai2009,Pillai2011,Gravel2011}, which combine trophic and geographical complexity. Meta-foodwebs describe the interactions between several different food webs in space and one particular class of meta-foodwebs is described by the colonization-extinction model proposed by Pillai \cite{Pillai2009,Pillai2011}. 

Meta-foodwebs can be described as networks of networks or multilayer networks \cite{Kivela2014,Boccaletti2014}
. To connect the ecological system to physical terminology one can regard the system from two different perspectives. The first of these focusses on the food web: We can say that meta-foodwebs are collections of food-webs that exist in different spatial patches and interact through the dispersal of individuals between patches. Seen from this perspective, the food webs are the layers of the network, predator-prey interactions are within-layer interactions, whereas dispersal of individuals between patches constitutes between layer interactions. 

We can describe the same class of systems in a different way by saying that meta-foodwebs are geographical networks of species dispersal that interact through feeding interactions. Now the network layers are formed by the geographical network, dispersal between patches is a within-layer interaction, whereas the feeding interactions constitute between layer interactions. 

The former perspective is useful when the food web is more complex than the geographical network, whereas the latter is useful when the geographical network is more complex than the food web. In ecology plenty of examples for both cases are encountered, and thus it may be useful to apply the elegant notation proposed in \cite{DeDomenico2013}. However, in the present paper we focus on the case where the food web is very simple (a linear chain), whereas the geographical network is both larger and complex in structure, we therefore employ the later perspective. We thus regard the spatial dispersal networks as network layers, which interact through feeding interactions.

Pillai's meta-foodweb model has been studied in the ecological literature using agent-based simulations. The central ecological question driving this work is how landscape structure impacts food web structure\cite{Pillai2009,Pillai2011}. It was shown that higher connectivity in the geographical networks generally benefits the persistence of species on the landscape level. However, in very strongly connected systems, specialist species tend to outcompete generalist  species, such that most complex food webs are found at intermediate geographical connectivity\cite{Pillai2011}.  

A theoretical approach for the computation of persistence thresholds in colonization-extinction models has been proposed in \cite{Bohme2012}. This work used the so called \emph{homogeneous approximation} in which all patches are considered to have the same number of links (degree). However, previous work on epidemics has demonstrated, that spreading processes can be understood well by utilising the power of generating functions\cite{Newman2001,Newman2002}. In particular, such approaches can be used to reveal the degree distribution (the probability distribution of links per node) of nodes in a particular state.

By using generating functions we can find the degree distribution of the network of geographical patches experienced by each species. These degree distributions differ from the degree distribution of the underlying geographical network, because some patches may be inhospitable to a given species due to its interactions with other species. For instance the absence of suitable prey can make a patch inhospitable to a given predator and thus removes the node from the network accessible to a predator. By revealing the degree distributions experienced by the different species, generating functions hold the promise of enabling a deeper understanding of the impact of landscape connectivity on ecological dynamics. 

Here we present a generating function approach to analyse how the degree distribution of the patch network affects food chains,  in which each species has at most one predator and one prey. We find that properties such as the shape and mean of the patch degree distribution affects the occupation probability of all species in the food chain, and the viability of survival for the species at the top of the food chain. Beyond the ecological insights, this paper highlights meta-foodchains and meta-foodwebs as promising example systems for the future refinement of tools of statistical physics.

\section{The Model}

We study a version of the model proposed by Pillai et al. in  \cite{Pillai2009}. A set of species numbered $1$ to $s$ inhabits an environment comprising a set of discrete patches. This environment is represented by a network, where nodes represent the patches and links represent the possible routes of dispersal between patches.  The model accounts for the presence, or absence, of each species in the food chain at each patch. 

The populations of different species interact with each other via trophic (feeding) relationships. Species $1$ is a so-called primary producer, a species that can persist on abiotic resources. In the model this species can colonize any patch independently of the presence of other species. All other species, $i>1$, are specialist consumers who each prey upon a single species, $i-1$. Therefore, a species $i$ can only inhabit a patch that species $i-1$ also inhabits. 

The system varies dynamically due to random extinctions of species at individual patches and colonisations of patches by new species. When established at a patch, species $i$ is subject to random extinction at rate $e_i$.  The interaction between species means that when species $i$ goes extinct on a patch, all species $j>i$ must also go extinct at that patch, because they now lack an essential resource farther down the chain.  This indirect extinction means that a species $i$ will go extinct on a particular patch at an effective rate equal to the sum of its direct extinction rate, $e_i$, and the extinction rates of all species below it in the food chain. When established at a patch species $i$ may also colonise neighbouring patches at a rate, $c_i$, however due to the trophic interactions a species $i$ can only colonise a patch at which its prey, species $i-1$, is already established.

While indirect extinction is a process between nodes of a food chain at a single patch, colonisation is a process between different food webs. Despite this difference both imply that for the dynamics of a given species $i$ only the subnetwork of patches where species $i-1$ is established is relevant. We call this network the effective network for species $i$.

Over time the effective network for any species $i>1$ changes due to colonization and extinction events of species $i-1$. These events affect $i$'s effective network by changing its size, the number of nodes; its connectivity, the number of links per node; and its degree distribution, the probability distribution of the number of links per node. Here we present a method for finding the effective network of all species $i>1$ and hence the pattern of dispersal for all species in the food chain.

\section{Abundance of Primary Producer}
To study the dispersal of species we describe the system on the level of the configuration model, where a given network is characterised by its degree distribution. 

\begin{figure}[t]
\centering  
\includegraphics{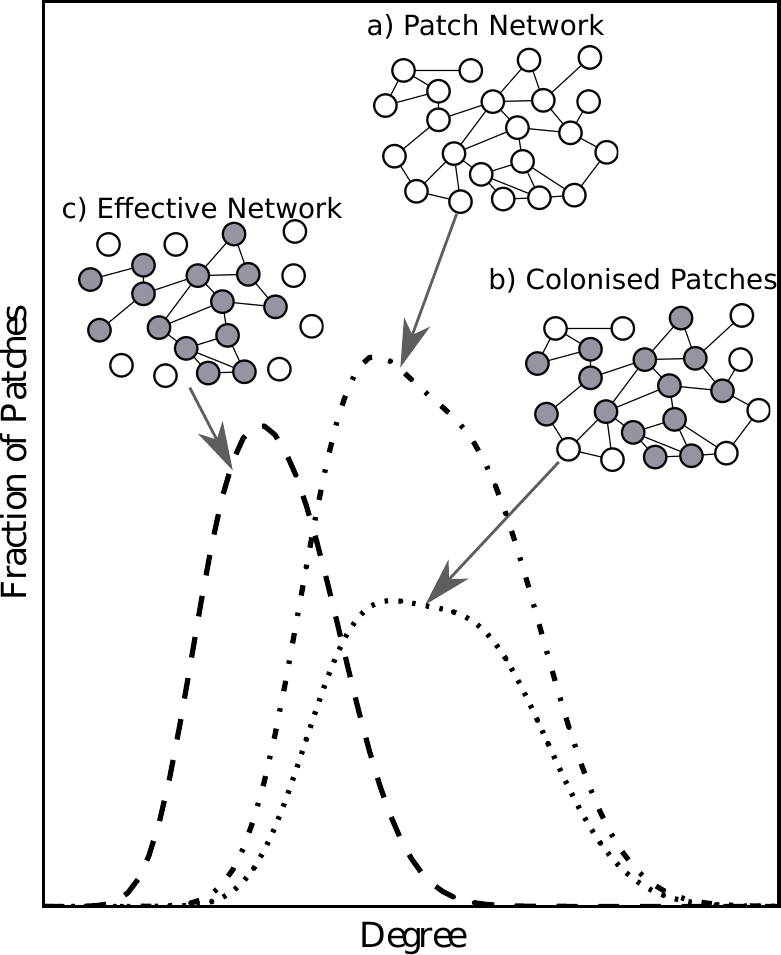}
\caption{ \label{DisChange} Example of the two step process for finding the degree distribution of the effective network of species $2$. We start with the patch network and its degree distribution, $p_k$, (a).  The first step identifies the nodes colonised by the species, $\chi_k$, (b). The second step removes links from colonised nodes to uncolonised nodes from the distribution giving the effective network, $g_k$, (c). This is the network upon which species $2$ disperses. }
\end{figure}

From the degree distribution of the patch network we consider the dispersal of species $1$ to find the degree distribution of the effective network for species $2$ via two steps (Fig. \ref{DisChange}):
\begin{enumerate}
\item Finding the expected degree distribution of patches in which species $1$ is established.
\item Removing links  from this distribution which lead to patches in which species $1$ is not established, and which are hence inaccessible to species $2$.
\end{enumerate}

Similarly we use the same two steps to consider the dispersal of species $2$ over the degree distribution of its effective network to find the effective network of species $3$.  By repeating this process through successive levels we find the degree distribution of all effective networks, and thus the properties of dispersal for all species $1$ to $s$.

To find the expected degree distribution of the patches inhabited by species $1$ we find the degree-dependent probability that species $1$ inhabits a patch with degree $k$. The probability that a patch has degree $k$ is $p_k$. Furthermore, we define $\chi_k$ to be the probability that a patch has both degree $k$ and is inhabited by species $1$ (Fig. \ref{DisChange}). 

In contrasts to $p_k$, $\chi_k$ is a dynamical variable which changes in time due to colonization and extinction events. For a sufficieantly large system these dynamics can be captured by the differential equation 
\begin{align}\label{Dchi}
\frac{\d\chi_k}{\d t}=-e_1\chi_k+c_1k(p_k-\chi_k)P_L(C|E),
\end{align}
where the first term captures random extinctions and the second term captures the effect of colonisations. In the colonisation term $P_L(C|E)$ is the probability that a link with an empty patch at one end has a colonised patch at the other. If we assume no correlation between the state of neighbouring patches then $P_L(C|E)=\sum_j j\chi_j/\sum_j jp_j$. This overestimates the potential for colonisation and it is shown in \cite{Shrestha2015} that for a significant parameter region a better approximation can be obtained by correcting for backtracking, which yields $P_{B}(C|E)= \sum_j (j-1)\chi_j/(\sum_j kp_j-\chi_j)$.

Systems of the form of Eq.~(\ref{Dchi}) typically either approach a steady state where $\chi_k=0$ for all $k$, such that the species goes extinct, or a nontrivial steady state in which the population survives. To compute this nontrivial steady state we must solve the system of differential equations for all patch degrees. Here we follow the approach of \cite{Silk2014} and use the method of generating functions to transform the system of equations into a single partial differential equation, which can then be solved for the desired steady state.

We encode the degree distribution of the patches by the probability generating function, $P(x)=\sum_k p_k x^k$ and the dynamically chaining probability generating function for the colonised patches $C(x)=\sum_k \chi_k x^k$. We can now write 
\begin{align*}
\frac{\rm d}{{\rm d}t} C(x) = \sum_k  x^k \frac{{\rm d}\chi_k}{{\rm d}t}
\end{align*}
which with Eq.~(\ref{Dchi}) yields
\begin{align}
\frac{\d{C}(x)}{\d t}&=\sum_k\bigg(-e_1\chi_k+c_1k(p_k-\chi_k)\frac{C'(1)-C(1)}{P'(1)-C(1)}\bigg)x^k\nonumber\\
&=c_1 x (P'(x)-C'(x))\frac{C'(1)-C(1)}{P'(1)-C(1)} - e_1 C(x)\label{CPDE},
\end{align}
where we used the dash to indicate the derivative with respect to $x$. These derivatives appear as we use the common trick \cite{Silk2014} of shifting the summation index by moving factors of $x$ outside of sums to create expressions that can be written as $P(x)$, $C(x)$ or their derivatives $P'(x)$, $C'(x)$.

Setting the left hand side of Eq.~(\ref{CPDE})  to zero we find the steady state condition
\begin{align}
C'(x)=-\frac{\gamma_1}{\alpha_1}\frac{C(x)}{x}+P'(x)\label{C1},
\end{align}
where we have let $\alpha_1=[C('1)-C(1)]/[P'(1)-C(1)]$ and $\gamma_1=e_1/c_1$ . Integrating gives
\begin{align}
C(x)=\bigg(\int P'(x) x^{\frac{\gamma_1}{\alpha_1}}\d{x}+a_I\bigg)x^{-\frac{\gamma_1}{\alpha_1}}\label{sol1},
\end{align}
where $a_I$ is a constant of integration. 

We determine $\alpha_1$ by setting $x=1$ in Eq.~(\ref{C1}),
\begin{align}
\alpha_1=\frac{\gamma_1C(1)}{P'(1)-C'(1)}\label{alpha1},
\end{align}
We can use Eq.~(\ref{sol1}) and Eq.~(\ref{alpha1}) to find $C(x)$ and $\alpha_1$, from which the whole distribution $\chi_i$ can be recovered by a Taylor expansion. For a given degree distribution this is most easily done numerically though there are some insights we can offer analytically. 

From the power series definition of $P(x)$ we can solve the integral explicitly for the general case to find the occupied fraction of nodes of each degree, the values of $\chi_k$. This integration gives 
\begin{align*}
C(x)=\sum_k \chi_k x^k &=\left(\int \sum_k k p_k x^{k-1} x^{\frac{\gamma_1}{\alpha_1}}\d{x}+a_I\right)x^{-\frac{\gamma_1}{\alpha_1}}\\
&=\left(\mathlarger{\sum_k} \frac {k p_k}{k+\frac{\gamma_1}{\alpha_1}} x^k x^{\frac{\gamma_1}{\alpha_1}}+a_I\right)x^{-\frac{\gamma_1}{\alpha_1}}\\
&=\mathlarger{\sum_k} \frac {k p_k}{k+\frac{\gamma_1}{\alpha_1}} x^k +a_I x^{\frac{-\gamma_1}{\alpha_1}}
\end{align*}
and by comparison of the coefficients of the polynomial we find the occupation probability of a node of degree $k$ is $k(k+\gamma_1/\alpha_1)^{-1}$.
\begin{figure*} [ht] 
\centering  
\includegraphics{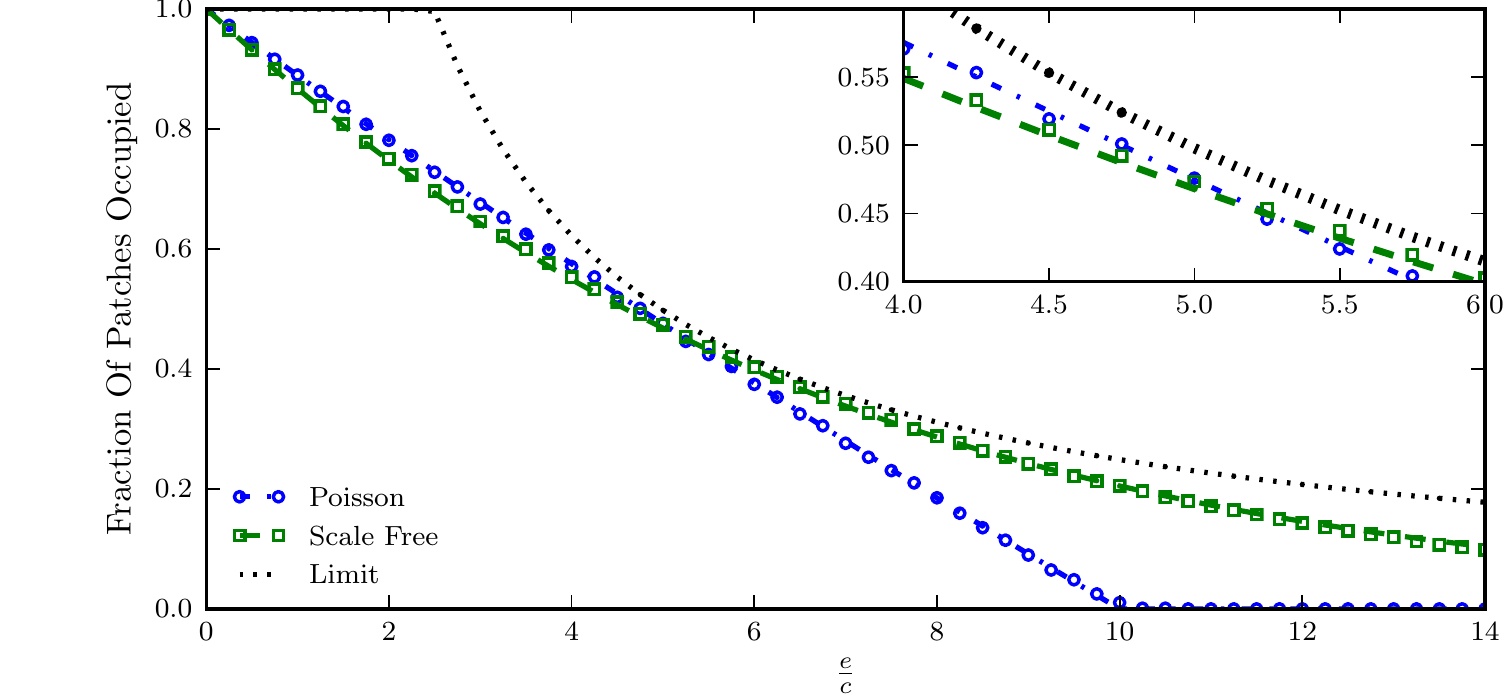}
\caption{\label{FractionsFig}(Colour online) Comparison of colonised abundance of species $1$ in patch networks with different degree distributions. Shown are analytical (lines) and simulated (markers) abundances for species on patch networks with poisson (blue) and scale-free (green) degree distributions with the same mean degree and the limit as calculated from the configuration model. For low $e/c$ the  poisson degree distribution leads to greater abundance but for high $e/c\gtrapprox 5$ it is the scale-free distributions that allows greater abundance. The insert highlights the region at which the lines cross.} 
\end{figure*}
We find an upper limit for the fraction of patches inhabited by using the approximation $P_L(C|E)=\sum_j j\chi_j/\sum_j jp_j$. As this overestimates the fraction the limit applies despite the approximation. For this case we have an expression similar to Eq.~(\ref{C1}) but with $\alpha_L=C'(1)/P'(1)$,
\begin{align}
C'(x)=-\frac{\gamma_1}{\alpha_L}\frac{C(x)}{x}+P'(x),
\end{align}
setting $x=1$ in this equation gives
\begin{align}
\alpha_L=1-\frac{\gamma_L}{\alpha_L}\frac{C(1)}{P'(1)}\nonumber \\
\alpha_L^2-\alpha_L+\gamma_1\frac{C(1)}{P'(1)}=0\label{alphaL}.
\end{align}
Solving for $\alpha_L$ using the quadratic formula we find
\begin{align}
\alpha_L=\frac{1}{2}\left(1\pm\sqrt{1-4\gamma_1\frac{C(1)}{P'(1)}}\right).
\end{align}
Given that $\alpha_L$ must be a real number we require $C(1)\leq P'(1)/(4\gamma_1)$. For a given mean degree $P'(1)$ and for given $\gamma_1$ there is a maximum possible inhabited fraction for any degree distribution ie.~with free choice of degree distribution the expected inhabited fraction will never be larger than this limit. 

To observe some of the effect of patch degree distribution on the abundance (fraction of patches inahbited) we compared the  abundances of species $1$ for patch degree distributions from two different random network models and for a range of values of $\gamma_1$.  The two models generate patch networks with very different degree distributions. One is the \ER model that generates a network with a poisson degree distribution  $p_k=\mean{k}^k\mathrm{e}^{-\mean{k}}/k!$ \cite{Gilbert1959,erdds1959random,erdos1961evolution}. The other Barab\'asi-Albert model which generates a network with a scale-free degree distribution, $p_k\propto k^{-3}$ \cite{barabasi1999emergence}. 

Using Eq.~(\ref{sol1}) and Eq.~(\ref{alpha1}) we found the theoretical abundance of species $1$ for different values of $\gamma_1$ for the average degree distributions of networks with $10^5$ nodes and mean degree $\left<k\right> = 10$ generated by each of the random network models. We also simulated the dispersal of species on the networks generated by each of the models for each value of $\gamma_1$ using a Gillespie algorithm \cite{Gillespie1976} from an initial state where half the nodes are inhabited.

We found that the distribution which enables greater abundance of species $1$ is dependent on $\gamma_1$ (Fig. \ref{FractionsFig}).  For low $\gamma_1$ the abundance of species $1$ is greater for the poisson distribution, where as for higher $\gamma_1$, when it is harder for the species to disperse, abundance is greater for the scale-free distribution. 

It is well known that finite scale-free networks have much higher epidemic thresholds than \ER random graphs with the same mean degree \cite{Pastor-Satorras2001}. Correspondingly we find that species $1$ survives on the scale-free distribution for values of $\gamma_1$ well above values for which it is extinct on the poisson distribution.

\section{Effective networks of predators}
\begin{figure}[b]
\centering  
\includegraphics{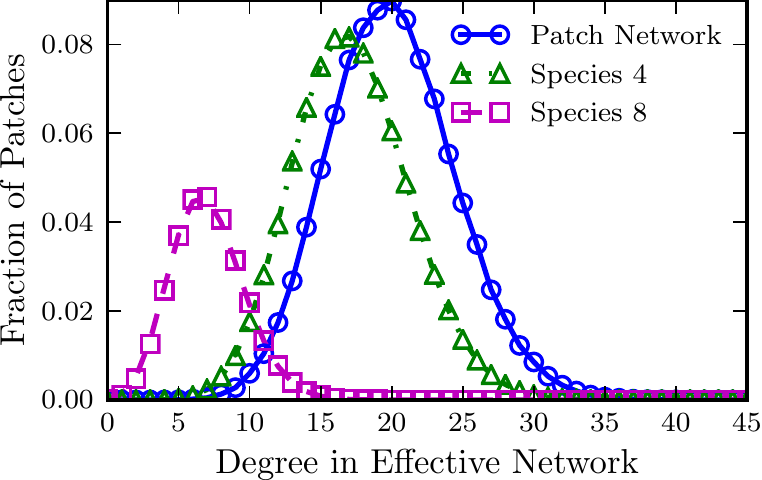}
\caption{\label{ERExample}(Colour online) Effective degree distributions experienced by species on different trophic levels for poisson patch networks with $\left<k\right>=20$. Shown are the degree distribution of a particular patch network and the effective network of species $4$ and $8$ food chain with which all species $e_i/c_i=0.5$ from simulation (markers) and analytically (lines). The sum of the distribution is reduced for higher trophic levels as the species inhabits fewer patches.} 
\end{figure}

\begin{figure} [b] 
\centering  
\includegraphics{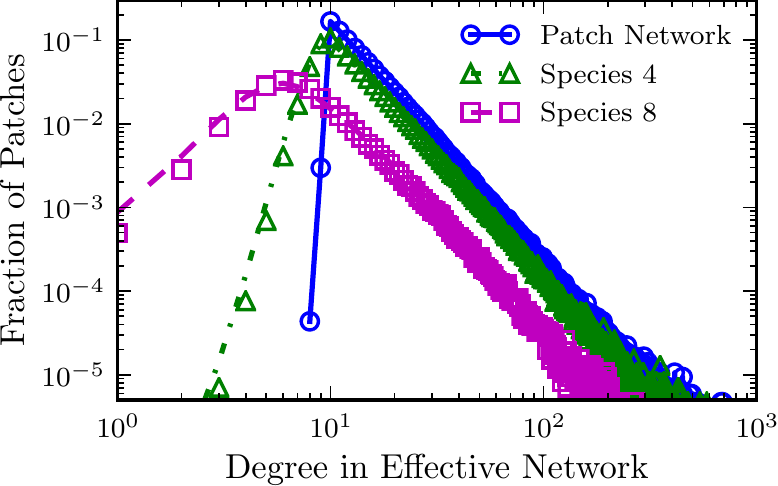}
\caption{ \label{BAExample}(Colour online) Effective degree distributions experienced by species on different trophic levels for scale-free patch networks with $\left<k\right>=20$. Shown are the degree distribution of a particular patch network and the effective network of species $4$ and $8$ food chain with which all species $e_i/c_i=0.5$ from simulation (markers) and analytically (lines). The sum of the distribution is reduced for higher trophic levels as the species inhabits fewer patches.}
\end{figure}

We now determine the effective network on which species $2$ is spreading from the degree distribution of nodes inhabited by species $1$, $C(x)$. A node with $k$ links in the patch network which is inhabited counts towards the $k^{\mathrm{th}}$ element of the distribution $C(x)$. For the effective network of species $2$ we are interested in only the links connecting nodes which are both inhabited by species $1$. We want an inhabited node of degree $k$ in the patch network which has $l$ links to other inhabited node to count towards the $l^{\mathrm{th}}$ component of the distribution of the effective network and we denote the function that generates this distribution, $G_2(x)$. 

To find $G_2(x)$ we remove the links from the degree distribution $C(x)$ which connect inhabited and empty nodes. Using the approximation $P_{B}(C|E)=[C'(1)-C(1)]/[P'(1)-C(1)]$, the probability of a link which has a colonised patch at one end having an empty patch at the other is
\begin{align}
P_{B}(E|C):=\beta_1&=\frac{\bigg[C'(1)-C(1)\bigg]\bigg[1-\frac{C'(1)}{P'(1)}\bigg]}{\bigg[P'(1)-C(1)\bigg]\frac{C'(1)}{P'(1)}}\\
&=\frac{\bigg[C'(1)-C(1)\bigg]\bigg[\frac{P'(1)}{C'(1)}-1\bigg]}{P'(1)-C(1)}.
\end{align}
We define $R_1(x)=\beta_1+(1-\beta_1)x$, a function that generates the distribution $r_0=\beta_1$ and $r_1=1-\beta_1$, ie. the distribution of whether a link from a colonised node ends at an inhabited node.

To find $G_2(x)$ we consider each link in $C(x)$ to exist with probability $\beta_1$. Therefore $G_2(x)=C(R(x))$ and so
 \begin{align}
G_2(x)=\bigg(\alpha_1\int P'(R_1(x)) (R_1(x))^{\frac{\gamma_1}{\alpha_1}}\d{x}\bigg)(R_1(x))^{\frac{-\gamma_1}{\alpha_1}}.\label{intG1}
\end{align}
As $C(1)=G_2(1)$ and $C'(1)=G_2'(1)/\beta_1$ Eq.~(\ref{intG1}) and Eq.~(\ref{alpha1}) can be solved together to find the degree distribution of the effective network for species $2$ directly from the degree distribution of the patch network.

We now have all the tools we need to find the effective network of all species. Species~$i$ disperses on its effective network, $G_i(x)$ with the relative extinction by colonisation rate $\gamma_i=\sum_{j \leq i} e_j /c_i$. Hence the degree distribution of the effective network of species $i+1$ is given by
\begin{align}
G_{i+1}(x)=\bigg(\alpha_i\int G_i'(R_i(x)) (R_i(x))^{\frac{\gamma_i}{\alpha_i}}\d{x}\bigg)(R_i(x))^{\frac{-\gamma_i}{\alpha_i}}\label{Gi}
\end{align}
with
\begin{align}
\alpha_i=\frac{\gamma_iG_i(1)}{G'_{i-1}(1)-\frac{G_i'(1)}{\beta_i}}.\label{alphai}
\end{align}

We find that there is good agreement between theoretical results from Eq.~\ref{Gi} and Eq.~\ref{alphai} with results and simulations for the degree distributions of effective networks for species at various trophic levels for patch networks with both poisson (Fig.~\ref{ERExample}) and scale-free (Fig.~\ref{BAExample}) degree distributions. 

The peak of the distribution moves left for successive effective networks, indicating the preferential inhabitation of high degree nodes is counteracted by a larger effect of the removal of links to empty nodes. Therefore the mean degree of the distribution decreases for the effective network of successive levels and thus dispersal is more difficult for species higher in the food chain. For the poisson patch degree distribution we note that the excess mean degree decreases faster than mean degree and therefore the effective networks do not have poisson degree distributions.

\section{Abundance of predators}

\begin{figure} [t] 
\centering  
\includegraphics{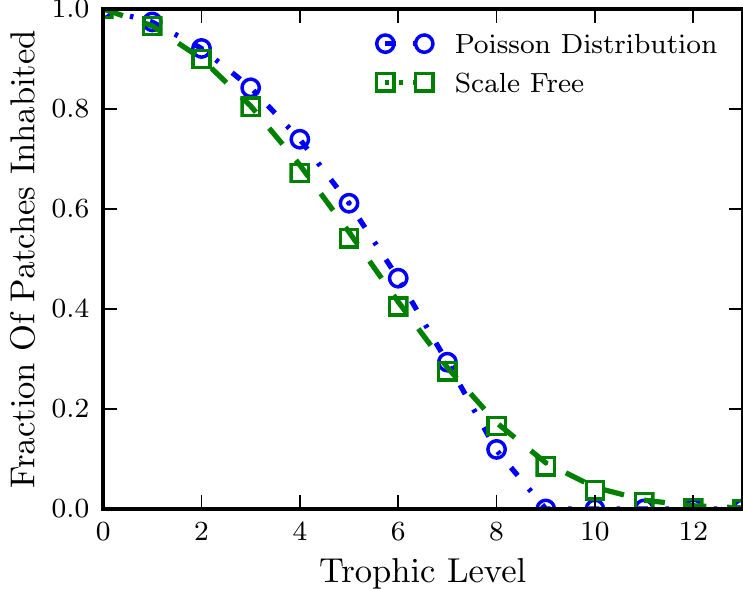}
\caption{\label{LevelsFig}(Colour online) Colonised abundance for many trophic levels of a food chain. Shown are the fraction of patches inhabited by each species in food chains on patch networks with poisson degree distributions (blue)  and scale-free degree distributions (green). All species have $e/c=0.5$ and both patch networks have $\left<k\right>=10$. For species in the low trophic levels the networks with poisson degree distribution have greater abundance but for species higher in the food chain it is the scale-free distribution that has the greater abundance. Further the scale-free network can support more species that the poisson network.} 
\end{figure}

The theoretical abundance of species $i$ is given by $G_{i+1}(1)$, which we find by setting $x=1$ in Eq.~\ref{Gi}. Comparing the abundance of species at different levels for a particular value of $\gamma_i$ we find that the patch network which gives the largest abundance of species low in the food chain is not necessarily the one that gives the greater abundance for species high in the food chain, fig.~\ref{LevelsFig}. The poisson network has the highest abundance of species low in the food chain however the scale-free network has greater abundance of species high in the food chain. Further the scale-free network can support more species than the poisson network. 

For infinite scale-free networks with $p_k \propto k^{-3}$ it is know that there is no epidemic limit and a species will survive for all values of $\gamma$ \cite{Pastor-Satorras2001}. This is due to the infinite variance of the degree distribution. We find that the variance of the effective degree distribution of species $i$ is 
\begin{align*}
\mathrm{Var}\left(g^{(i)}_k\right)=\gamma_i G_i(1)+\alpha_i^2G_{i-1}''(1)+\alpha_i  G_{i-1}'(1)-(G_i(x))^2,
\end{align*}
from which we see that if the patch degree distribution has infinite variance, as $P''(1)$ is infinite, then the variance of all effective degree distributions will also be infinite. Therefore we expect that there is no epidemic limit for all species in a food chain on an infinite scale-free network.

\section{Conclusions}

We have presented a mathematical approach to the degree distribution of the network accessible to species at various levels of a food chain. For the finite patch networks we found  the network accessible both shrinks and becomes harder to spread over for species at successive levels. Hence the maximum effective extinction rate with which a species can survive decreases as we consider species higher in the food chain.

The analytical solutions also indicate a maximum abundance for a species, $G_{i-1}'(1)/(4\gamma_i)$, which is dependent on the effective extinction rate, $\gamma_i$, and the mean degree of the network it disperses over,$G_{i-1}'(1)$. Our results indicate that there is no degree distribution that is advantageous at all values of $\gamma_i$ for a particular mean degree. Importantly, we found that while very heterogeneous, scale-free, distributions come very close to the theoretical optimum at high $\gamma_i$, less heterogeneous distributions lead to higher abundances when $\gamma_i$ is lower.

One implication of the results is that species close to the bottom of the food chain can sometimes profit from homogeneous degree distributions. However, species higher up in the same food chain may nevertheless be more abundant if the underlying topology is more heterogeneous. That means a finite scale-free distribution is likely to allow more species to survive than a poisson distribution even when the latter allows a greater abundance of species low in the food chain. 

In land use planning one is often forced to decide which patches of forest to conserve, or where to place green spaces in a city. Hence the planner has some control over the degree distribution of the patch networks these created by these processes. If seeking to maximise the abundance of a particular species our results show that a good estimation of $\gamma_i$ is required to inform any decisions. In addition maximising the abundance of the lowest species in the food chain may not maximise the number of species that survive. This is of particular note in the real world where small populations may be less resilient to external shocks.

The approach we use provides the tools required to analyse the behaviour of food chains on patch networks with varying degree distributions. Investigations into degree distributions other than those presented here will provide more information about the implications of landscape distributions on the resident species.

In this paper we have focused solely on food chains, an important class of food webs that is at the focus of many ecological studies. However, many more complex food webs topologies also play a significant role in ecology \cite{Gross2009}. In contrast to chains these webs also contain inter-species competition for prey, and predation on multiple prey species. This paper has established two operations in the algebra of Pillai-style colonization extinction models. The pruning of links to patches where an essential resource is missing and the subsequent renormalization of the degree distribution. To accommodate additional interaction that occur in food webs an additional operation is necessary, the pruning of patches where a superior competitor is established. This operation is similar to the pruning of patches studied here and, although the notation does get more cumbersome, no fundamental obstacles should arise in this step. Thus the approach proposed here can be adapted to deal with more complex webs.

The treatment presented here was relatively easy because we were able to compute the degree distributions iteratively, from the bottom layer up. The same is also true for some more complex food web topologies as long as their is a clear hierarchy in the strength of competitors. Some ecological interactions break these hierarchies. A simple example is neutral competition where two competitors can defend a patch in which they are established against the other species. A mathematically more interesting scenario (which is fortunately rare in ecology) is the case where predators can drive their own prey to extinction. In both cases interdependencies arise such that systems of generating functions have to be solved simultaneously. Dealing with these non-iterative cases is mathematically more difficult, but could also lead to much richer dynamics. They thus present a promising target for future work.

\begin{acknowledgements}
This work was completed in close collaboration with DFG Research Unit 1748 (Networks on Networks) and supported partially by the EPSRC under grant code EP/I013717/1(BCCS).
\end{acknowledgements}

\bibliographystyle{ieeetr}

\end{document}